# Signature splitting inversion and backbending in $^{80}$Rb

Chuangye He[1]  Shuifa Shen[2 3 4, *]  Shuxian Wen[1]  Lihua Zhu[5]  Xiaoguang Wu[1]  Guangsheng Li[1]  Yue Zhao[3]

Yupeng Yan[2 6]  Zhijun Bai[4]  Yican Wu[7]  Yazhou Li[7]  Gui Li[7]  Shiwei Yan[8]  M. Oshima[9]  Y. Toh[9]  A. Osa[9]

M. Koizumi[9]  Y. Hatsukawa[9]  M. Matsuda[9]  T. Hagakawa[9]

1 China Institute of Atomic Energy, P. O. Box 275(10), Beijing 102413, China

2 School of Physics, Suranaree University of Technology, Nakhon Ratchasima 30000, Thailand

3 State Key Laboratory Breeding Base of Nuclear Resources and Environment, East China Institute of Technology, Nanchang 330013, Jiangxi, China

4 State Key Laboratory of Nuclear Physics and Technology (Peking University), Beijing 100871, China

5 School of Physics and Nuclear Energy Engineering, Beihang University, Beijing 100191, China

6 Thailand Center of Excellence in Physics (ThEP), Commission on Higher Education, 328 Si Ayutthaya Road, Ratchathewi, Bangkok 10400, Thailand

7 Institute of Nuclear Energy Safety Technology, Chinese Academy of Sciences, Hefei 230031, China

8 College of Nuclear Science and Technology, Beijing Normal University, Beijing 100875, China

9 Japan Atomic Energy Research Institute, Tokai, Ibaraki 319-1195, Japan

**Abstract**: High−spin states of $^{80}$Rb are studied via the fusion-evaporation reactions $^{65}$Cu+$^{19}$F, $^{66}$Zn+$^{18}$O and $^{68}$Zn+$^{16}$O with the beam energies of 75 MeV, 76 MeV and 80 MeV, respectively. Twenty-three new states with twenty-eight new γ transitions were added to the previously proposed level scheme, where the second negative-parity band is significantly pushed up to spins of 22$^-$ and 15$^-$ and two new sidebands are built on the known first negative-parity band. Two successive band crossings with frequencies 0.51 MeV and 0.61 MeV in the α=0 branch as well as another one in the α=1 branch of the second negative-parity band are observed for the first time. Signature inversions occur in the positive- and first negative-parity bands at the spins of 11ℏ and 15ℏ, respectively. The signature splitting is seen obviously in the second negative-parity band, but the signature inversion is not observed. It is also found that the structure of the two negative-parity bands is similar to that of its isotone $^{82}$Y. Signature inversion in the positive-parity yrast band with configuration $\pi g_{9/2} \otimes \nu g_{9/2}$ in this nucleus is discussed using the projected shell model (PSM).

---

*E-mail address: shfshen@ecit.cn

## I. INTRODUCTION

The character of the nuclei in the mass A=80 region shows a strong competition between the single particle excitation and collective rotation. Existing research results indicate that nuclei with the neutron number less than 44 possess strong collective rotations while those with the neutron number more than 47 show obviously single particle excitations. In the end of last century, Neutron−deficient odd−odd nuclei in the A=80 mass region attracted so large attentions that a great many of experimental research results have been achieved in nuclei, e.g., $^{76}$Rb[1], $^{78}$Rb[2], $^{78}$Br[3], and $^{82}$Y[4] etc. Our present experimental research object is focused on $^{80}$Rb which has 43 neutrons and hence its structure is mainly characterized by the collective rotation.

So far, in the neutron deficient nucleus $^{80}$Rb high-spin states have been studied via the reaction $^{51}$V($^{32}$S, 2pn)[5], and the spins of the positive-parity yrast band and first negative-parity band were extended to 25ℏ and 23ℏ, respectively. Later in 2000, the high−spin states of $^{80}$Rb were populated through the $^{55}$Mn($^{28}$Si, 2pn) reaction at 90 MeV and the quadrupole deformations |β$_2$|≈0.3, which are predicted to be nearly oblate, of 13 levels and the lower limits for 4 levels in $^{80}$Rb were determined experimentally[6]. In addition, the total Routhian surface (TRS) calculations carried out by Cardona *et al*.[6] predict that for the positive-parity states with low frequencies, the nucleus $^{80}$Rb is γ soft, with a quadrupole deformation β$_2$≈0.33. As the frequency increases, for example at ℏω=0.492 MeV, two minima become visible at γ=19° and

γ=−30°. The second minimum remains over the entire range of frequencies measured in that experiment, i.e., for spins greater than 9ℏ, and evolves towards an oblate shape as the frequency increases. The TRS calculations for the negative-parity states of the π($p_{1/2}$ or $p_{3/2}$ or $f_{5/2}$)⊗ν$g_{9/2}$ configuration predict an oblate equilibrium shape with similar deformation parameters over the entire range of measured frequencies, i.e., above the I=$9^-$ level. Our purpose of the present study is to extend the levels to higher spins, especially populate more side bands with lighter projectile and study its structural features. The paper is arranged as follows: the experimental details and results are given in Section II. We analyze and discuss in Section III experimental results of the signature splitting and inversion of the negative-parity band of $^{80}$Rb as well as its neighboring isotones and isotopes. Theoretical study of the positive-parity yrast band is presented in detail in Section IV. Finally the work is summarized in Section V.

## II. EXPERIMENTAL DETAILS AND RESULTS

High−spin states in $^{80}$Rb were populated via fusion-evaporation reactions using the 75 MeV $^{19}$F, 76 MeV $^{18}$O and 80 MeV $^{16}$O beams (The intensities of these three beams are all $I$∼1pnA) provided by the HI−13 Tandem accelerator at China Institute of Atomic Energy (CIAE) and Tandem accelerator at Japan Atomic Energy Research Institute (JAERI), respectively, and the beam energy was chosen on the basis of cross-section calculations and excitation function measurements. The γ − γ coincidence and directional correlation of oriented nuclei (DCO) ratios were measured by two detector arrays consisting of 10 (efficiencies 25% to 35%) and 12 (efficiencies 40% to 60%) Compton−suppressed HPGe−BGO detectors at CIAE and JAERI, respectively. Each detector had an energy resolution of about 2 keV for 1332.5 keV γ−ray. Energy and relative efficiency calibration of the detector was performed using standard sources such as $^{60}$Co and $^{152}$Eu mounted at the target position. In this experiment, The isotopically enriched $^{65}$Cu, $^{66}$Zn and $^{68}$Zn targets are all self−supporting thin targets, which are rolled into thickness of 0.56 mg/cm$^2$, 0.62

mg/cm$^2$, and 0.57 mg/cm$^2$, respectively. The targets consist of a stack of two foils, it can increase the yield as well as avoid the deterioration of γ−ray energy resolution caused by Doppler effect in the thick target. These detectors were placed at ±30°, ±60° and ±90° relative to the beam direction. Events have been collected, in an event-by-event mode, when at least two Compton-suppressed Ge detectors fired in coincidence. A total of 2.0×10$^8$ and 3.0×10$^8$ double− or higher−fold coincidence events were accumulated in each experiment at CIAE by using the first two reactions and JAERI, respectively. A two-dimensional γ−γ coincidence matrix was established with experimental data of the reactions $^{65}$Cu+$^{19}$F and $^{66}$Zn+$^{18}$O. After accurate gain matching for the experimental data acquired at JAERI, we have constructed the γ−γ coincidence matrix and DCO ratio matrix. To obtain the DCO ratios of the γ rays, the data were sorted off-line into an angle-related $E_\gamma$-$E_\gamma$ matrix by placing the events recorded in the detectors at 90° on the x axis, whereas the events recorded in the detectors at ±30° angles were on the y axis. DCO ratio matrix is used to assign multipolarity of γ transition, then it help assign the spin of the relevant level. The DCO ratios deduced for strong γ rays in $^{80}$Rb are listed in Table I. The γ−γ coincidence data were analyzed with the Radware software package[7] based on a Linux-PC system. After gating on the previously known γ transitions, besides the known γ transitions we have identified more than 30 new γ transitions which belong to this nuclide. As an example, the sum coincidence spectrum gated on the 268, 367 and 646 keV γ rays is shown in Fig. 1. Gated spectra are produced for each of the γ−rays assigned to $^{80}$Rb. Selected coincidence spectrum obtained by gating on the 472 keV transition is shown in Fig. 2(a), where the transitions in positive−parity yrast band are clearly seen, and spectra obtained by gating on the 915 and 244 keV transitions are shown in Fig. 2(b) and 2(c), respectively, where most of the newly discovered γ rays in negative-parity bands are shown. Twenty−three new states with twenty−eight new γ transitions have been added to the previously proposed level scheme, as shown in Fig. 3. Compared to previous works[5-6], the present experiment has increased by 4ℏ the even spin sequences (α=0) of both the positive-parity yrast and first negative-parity bands and established some sideband transitions. Two E2 transition strings which have not been found before are observed sitting on the 11⁻ and 13⁻ levels of the first negative-parity band. And the present work pushes the spin of the second negative-parity band from 12⁻ to 22⁻ ($\alpha$=0) and from 11⁻ to 15⁻ ($\alpha$=1). The $\gamma$–ray above the 6⁺ level can not be observed from the prompt coincidence spectrum gated on the $\gamma$–ray below 6⁺ level because the 6⁺ state is an isomer, so the relative intensities of the $\gamma$–rays can not be given consistently. In addition, the relative intensities of the $\gamma$–rays can not also be derived from the total-projection spectrum because of the interference from other reaction channels.

Tandel *et al.*[5] assumed the configuration of the first negative-parity band of $^{80}$Rb to be ($\pi f_{5/2} \otimes \nu g_{9/2}$) two quasi–particle (qp) state in which the first backbending at the spin 15ℏ is due to the alignment of a pair of $g_{9/2}$ protons and the second backbending due to a pair of $g_{9/2}$ neutrons. Shown in Fig. 4 is the experimental alignments of the second negative-parity bands in $^{80}$Rb and its neighboring isotone $^{82}$Y[4], in which the Harris parameters used for reference, taken from Ref. [5], are $J_0$=21ℏ$^2$MeV$^{-1}$ and $J_1$=0ℏ$^4$MeV$^{-3}$. Two successive band crossings with frequencies 0.51 and 0.61 MeV in $^{80}$Rb were deduced from this figure. The first band crossing is very close to that in $^{82}$Y, the difference is only 0.02 keV. The highly similarity of their alignment patterns indicates that the $^{80}$Rb and $^{82}$Y have the similar configurations for the second negative-parity band. As the configuration $\pi[f_{5/2}+p_{1/2}]\otimes\nu g_{9/2}$ has been assigned to $^{82}$Y, one may conclude that the backbending of $^{80}$Rb in the frequency 0.51 MeV is formed by the uncoupling alignment of a pair of $g_{9/2}$ protons.

There may be a band crossing with frequency 0.54 MeV in the $\alpha$=1 branch of the second negative-parity band, it is slightly higher than that in the $\alpha$=0 branch of this band. It is difficult for us to define the nature of these band crossings in the absence of experimental information, such as life-time and g-factor measurements as well as relevant theoretical calculations. However, the possible explanation could be presented in comparison to that of its isotone $^{82}$Y as mentioned above, the adjacent nucleus $^{79}$Kr[8] and the first negative-parity band of $^{80}$Rb. Tandel *et al.*[5] and Paul *et al.*[4] have tentatively assigned the configurations $\pi[f_{5/2}+p_{1/2}]\otimes\nu g_{9/2}$ or $\pi f_{5/2}\otimes\nu g_{9/2}$ to the second negative-parity band. This configuration for two quasi-particle states means that a g$_{9/2}$ neutron pair cannot be a part of the four quasi-particle band, since the first neutron crossing is blocked. The isotone nucleus $^{79}$Kr (N=43) has the similar oblate ground state deformation and the first neutron crossing is also blocked, therefore it is able to compare to that in $^{80}$Rb. The first band crossing at 0.55 MeV and the second upbend at frequency 0.75 MeV in the negative-parity band in $^{79}$Kr has been attributed to the alignments of a pair of g$_{9/2}$ proton and a pair of g$_{9/2}$ neutron, which is consistent with that of total Routhian surface calculations in $^{79}$Kr[8], the latter indicated that the first band crossing occurs at frequency 0.5 MeV due to a pair of g$_{9/2}$ proton alignment and the second crossing above 0.6 MeV is attributed to a pair of g$_{9/2}$ neutron alignment. The similarities in the crossing frequencies (0.51 and 0.61 MeV of $^{80}$Rb) and the nature of interaction of the quasi-particle bands in $^{79}$Kr and $^{80}$Rb, can suggest that the first and second band crossings in the second negative-parity band of $^{80}$Rb result from the alignments of a $\pi$g$_{9/2}$ and a $\nu$g$_{9/2}$ pair, respectively.

### III. SIGNATURE SPLITTING AND INVERSION

The signature splitting and inversion are best visualized by plotting the experimental quantity [E(I)−E(I−1)]/2I as a function of the spin *I* of the initial state. The plot of [E(I)−E(I−1)]/2I vs. *I* for the positive-parity yrast band observed in the present work is similar to that of Tandel *et al.*[5] There are obvious signature splitting and inversion in this band, where the energy difference E(I)−E(I−1) of odd−spin states is higher in low spin region while the energy difference of even−spin states is higher after the signature inverses at the spin I=11$\hbar$ and the magnitude of the signature splitting increases with spin.

In this section, we discuss mainly the signature splitting and inversion of the negative-parity bands of $^{80}$Rb as well as its neighboring N=43 odd−odd isotones $^{78}$Br[3] and $^{82}$Y[4] and odd−odd isotopes $^{78}$Rb[2] and $^{76}$Rb[1]. Plotted as functions of spin *I* in Fig. 5(a), 5(b) and 5(c) are the energy differences [E(I)−E(I−1)]/2I of the first negative-parity bands of $^{78}$Br[3], $^{80}$Rb and $^{82}$Y[4], respectively. One sees from Fig. 5(a) that the signature splitting exists in this band and the signature inversion occurs at the spin I=15. It is also noted in Fig. 5 that the features of the quantity [E(I)−E(I−1)]/2I are quite similar among these three nuclei. The energy difference E(I)−E(I−1) of the odd−spin states is larger than that of even−spin states for low spin states, but this pattern reverses to that the energy difference E(I)−E(I−1) in even−spin states is larger than that of odd−spin states after $^{78}$Br and $^{80}$Rb exhibit the signature inversions at spin 14ℏ and 15ℏ, respectively. It can also be seen from Fig. 5 that, with the proton number increasing, the signature inversion happens to delay. The signature inversion in $^{82}$Y can not be observed even up to the spin state I=18. The patterns of the signature splitting and inversion of the first negative-parity band of $^{78}$Br and $^{80}$Rb are similar to that of their positive-parity yrast bands.

Shown in Fig. 6(a), 6(b) and 6(c) are the quantities [E(I)−E(I−1)]/2I, as functions of spin I, of the first negative-parity bands of $^{80}$Rb, $^{78}$Rb and $^{76}$Rb, respectively. It can be seen from Fig. 6 that for the Z=37 odd−odd isotopes the magnitude of signature splitting doesn't change much with the increasing of the neutron number at lower spin states. But the signature splitting pattern of $^{80}$Rb is different from that of $^{78}$Rb and $^{76}$Rb, that is, the energy difference E(I)−E(I−1) of odd−spin states of $^{80}$Rb is larger than that of even−spin states in the low spin region. After passing the signature inversion point 15ℏ state, the level energy difference E(I)−E(I−1) of even−spin states of all these three nuclei is larger than that of odd−spin states and the splitting increases gradually with the spin at higher spin states. The reason behind the presence of this different pattern in $^{80}$Rb with its isotopes is still an open question. Some of the earlier works[9-11] have been done by Liu *et al*. in A~130 and A~160 mass region. They reassigned the spins in some nuclei which are not according with the systematics. One may argue whether the assignment of spins in $^{80}$Rb is correct.

There may exist a reversal in phase of signature splitting pattern between the spins 7ℏ and 8ℏ in $^{80}$Rb, $^{78}$Br, $^{78}$Rb and $^{76}$Rb. The signature inversion in this mass region has been discussed based on the particle plus−rotor approaches[3, 12], which attributes the inversion to a change from the excitation modes involving both the quasi-particle alignment and collective rotation at low spins to involving only the rotation at high spins. The inversion should occur after the highest value available from the intrinsic motion of two quasi−particles in the odd−odd nucleus has been reached. Tandel *et al.*[5] has assigned the configuration $\pi f_{5/2} \otimes \nu g_{9/2}$ to first negative-parity band and a similar configuration to second negative-parity band, the highest value 7ℏ can be obtained from those configurations, above which the rotational band can be generated by collective motion. Therefore, the inversion at about 7ℏ could be understood by means of two fully aligned quasi−particles with the $\pi f_{5/2} \otimes \nu g_{9/2}$ and similar configurations connected with collective motion of a system. For the signature inversion occurs at spin 15ℏ in the first negative-parity band, Tandel *et al.*[5] has given an explanation by means of that in $^{79}$Rb[13] the α=0 signature partner is favoured below 15ℏ which remains in an oblate shape with a large negative value of γ approaching −60° while the α=1 component would be driven towards the prolate axis with γ=0° or a small triaxial deformation with a positive γ value.

Due to among the isotones of $^{80}$Rb only in $^{82}$Y the second negative-parity rotational band has been observed, so that the signature splitting of that bands of only these two isotones is shown in Fig. 7 which looks quite similar and is also similar to their positive-parity yrast band and first negative-parity band except for a little difference in low spin states. That there is signature difference in the low spin state 6⁻ may lay on the complexity of depopulated γ transitions in low spin states. In well-defined collective rotation bands, the energy of each level follows the regularity of I(I+1), but the low spin states may not follow this rule well, they still do not really feed into this collective rotation band. So far, no signature inversions have been observed in second negative-parity rotational bands in these two nuclei.

# IV. THEORETICAL STUDY OF POSITIVE-PARITY YRAST BAND

In this work, to study the underlying mechanism of signature inversion in the positive-parity yrast band in mass A~80 region, we apply the projected shell model (PSM) to the representative nucleus $^{80}$Rb. The theoretical model is described in detail in Ref. [14]; here we focus on the application of the PSM to $^{80}$Rb. Prior to the present work, there has been no information concerning the mechanism of the signature inversion about the positive-parity yrast band in this nucleus. The signature inversion phenomenon in $^{74}$Br, $^{76,78}$Rb, and $^{80,82}$Y has been studied in detail via the projected shell model (PSM) approach[15], where the basis deformation $\varepsilon_2$ is separately fixed for each nucleus. However, the calculations in Ref. [15] fail to reproduce the signature inversion observed at low spins in these nuclei, which suggests that one may need to consider other mechanisms that could cause the inversion.

In the present work, the spin–orbit force parameters, $\kappa$ and $\mu$, appearing in the Nilsson potential are taken from the compilation in Ref. [16], which is a modified version of that given in Ref. [17] that has been fitted to the latest experimental data then. The parameters in Ref. [16] are supposed to be applicable over a sufficiently wide range of shells. The values of $\kappa$ and $\mu$ are different for different major shells ($N$ dependent). About ten years ago, based on available experimental data then, a new set of Nilsson parameters was proposed by Sun et al.[18] for proton–rich nuclei with proton (neutron) numbers 28≤$N(P)$≤40. Considering that the nucleus studied in the present work has a neutron number 43, we believe that this new set of parameters is not very suitable for the nucleus $^{80}$Rb, although the Nilsson parameter set proposed by Zhang et al.[16] was deduced for A≈120−140. The pairing gaps are calculated using the four−point formula[19], where the binding energies of the relevant nuclei, $B$, are taken from Ref. [20], and experimental data are adopted if only they can be supplied. We obtain $\Delta_p$=1.215 MeV and $\Delta_n$=1.1025 MeV. In the present calculations, we carefully choose the monopole pairing strength $G_M$ for each type of nucleons, which approximately reproduce the observed odd–even mass difference in the mass region (i.e. the pairing gaps $\Delta_n$ and $\Delta_p$ deduced from the BCS calculations reproduce the above mentioned $\Delta_n$ and $\Delta_p$, respectively). Finally, the quadrupole pairing strength $G_Q$ is assumed to be proportional to the monopole strength, $G_Q$=0.16$G_M$. In this work, the quadrupole deformation parameter is taken as $\varepsilon_2$=−0.2832 from Ref. [6] as we believe that this value is more valid since it is deduced from the experimental data. It should be pointed out here that the relationship between the deformation parameters $\varepsilon_2$ and $\beta_2$ is the same as that in Ref. [21]. If the first term is adopted only, then $\varepsilon_2 = -0.2832$ equals approximately to $\beta_2 = -0.3$. The hexadecapole deformation parameter $\varepsilon_4 = 0.067$ is taken from the compilation of Möller et al.[20] In our calculations, the configuration space is constructed by selecting the qp states close to the Fermi energy in the $N=4$ ($N=4$) major shell for neutrons (protons), i.e., all orbitals of the $g_{9/2}$ subshell and the $K=5/2$ orbital of the $d_{5/2}$ subshell (the $K=3/2, 5/2, 7/2, 9/2$ orbitals of the $g_{9/2}$ subshell) is selected, and forming multi–qp states from them. Comparison of the experimentally observed signature inversion in the positive–parity yrast levels of $^{80}$Rb with the prediction of the PSM is given in Fig. 8. The experimental data are taken from our present work. Because of the absence of experimental datum for the $7^+$ level, the experimental transition energies between levels $8^+$ and $7^+$ and between levels $7^+$ and $6^+$ are not given in Fig. 8. As can be seen from Fig. 8, the agreement between the calculation and experiment is quite satisfactory above $I \approx 9\hbar$. The energy splitting at higher spins is well reproduced, indicating that the important influence on the yrast band from the low-$K$ components of the $g_{9/2}$ valance neutrons and protons are correctly accounted for by the configuration mixing. However, the calculation does not reproduce the signature inversion observed at low spin in $^{80}$Rb.

To analyze the deformation for the positive-parity state of this nucleus in detail, total Routhian surface (TRS) calculations are carried out by means of the pairing-deformation-frequency self-consistent cranked shell model[22-23]. Samples of TRS with signature $\alpha=0$ are presented in Fig. 9 in the polar coordinate plane ($\beta_2, \gamma$) at specific rotational frequencies $\hbar\omega=0.0, 0.2, 0.4$, and $0.6$ MeV corresponding to $I \sim (0-15)\hbar$, and the energy difference between neighboring contours is 200 keV. According to our TRS calculations, at a low rotational frequency, this nucleus is predicted to be very $\gamma$ soft. This is not surprising at all because our wide survey of various regions of the nuclear Periodic Table shows that the signature inversion phenomenon only occurs in soft nuclei. With increasing frequency, the nucleus becomes slightly more rigid. In TRS calculations, at rotational frequency $\hbar\omega=0.0$

MeV, which corresponds to ground state $1^+$ in $^{80}$Rb, the quadrupole deformation parameter is at $\beta_2$=0.218 and $\gamma$=-60.791°, which indicates it is a deformed nucleus with oblate shape. When the $\hbar\omega$ increases, this absolute minimum persists up to highest rotational frequency at $\hbar\omega$=1.0 MeV corresponding to $I$~22$\hbar$ calculated in the present work.

## V. SUMMARY

The high−spin states of $^{80}$Rb have been studied by using $^{65}$Cu+$^{19}$F, $^{66}$Zn+$^{18}$O, and $^{68}$Zn+$^{16}$O reactions. Twenty-three new states with twenty-eight new γ transitions have been assigned to $^{80}$Rb. In the present work, we have increased by 4$\hbar$ the levels in the α=0 sequences of both the positive-parity yrast and first negative-parity bands. In addition, two sideband transition strings have been built on the first negative-parity band. Furthermore, we extend the spins to 22$\hbar$ and 15$\hbar$ respectively in the second negative-parity band in which two consecutive band crossings with frequencies of 0.51 and 0.61 MeV at the α=0 branch as well as another one with frequency of 0.54 MeV at the α=1 branch have been observed for the first time. The signature splitting exists in all three bands, of which the positive- and first negative-parity bands show signature inversions at 11$\hbar$ and 15$\hbar$, respectively.

We have also made a comparison of the patterns of signature splitting and inversion in $^{80}$Rb with that of its neighboring isotones and isotopes. It is found that in the first negative-parity bands, with regard to the three isotones $^{78}$Br, $^{80}$Rb and $^{82}$Y, the signature inversions delay with the increase of proton number; The signature splitting pattern of this band in $^{80}$Rb is different from that of its neighboring isotopes $^{78}$Rb and $^{76}$Rb. The signature splitting patterns of the second negative-parity band and the positive-/negative-parity yrast bands in $^{80}$Rb are similar each other and also similar to that of the second negative-parity band in $^{82}$Y.

Finally, the signature inversion occurred in positive-parity yrast band is theoretically studied in the framework of projected shell model.

## ACKNOWLEDGEMENTS

Thanks to the staff of Tandem accelerator and target made group at CIAE and JAERI for providing the support to this work. The project is supported by the Major State Basic Research Development Program in China under Contract No. 2007CB815003, the National Natural Science Foundation of China under Grant Nos. 11065001, 61067001, 11075214，10927507, 11175259 and 10975191, the Foundation of the Education Department of Jiangxi Province under Grant No. GJJ12372, and Suranaree University of Technology under contract No. 15/2553.

Figure captions:

Fig. 1. Sum coincidence spectrum of γ rays produced by gating on the lines of 268, 367 and 646 keV.

Fig. 2. Spectra of *γ*-rays gated on (a) 472 keV, (b) 915 keV, and (c) 244 keV transitions, respectively. The *γ*-rays marked with an asterisk are contaminations originated from reaction products [77, 79, 80]Kr and [77]Br.

Fig. 3. A partial level scheme of $^{80}$Rb proposed in the present work. The transition energies are given in keV. It should be noted that the $6^+$ state is a μs isomer, and therefore the transitions feeding this state do not have a prompt coincidence with those below it.

Fig. 4. Experimental alignments of the second negative-parity bands in $^{80}$Rb and its neighboring isotone $^{82}$Y.

Fig. 5. Comparison of signature splitting and inversion for the first negative-parity bands of $^{78}$Br, $^{80}$Rb, and $^{82}$Y.

Fig. 6. Similar to Fig. 5, but for the first negative-parity bands of $^{80}$Rb, $^{78}$Rb and $^{76}$Rb.

Fig. 7. Similar to Fig. 5, but for the second negative-parity bands in $^{80}$Rb and $^{82}$Y.

Fig. 8. Transition energies of the positive-parity yrast band of $^{80}$Rb as a function of the spin $I$ of the initial state. The energy difference E(I)-E(I-1) is compared between the theoretical predictions (open circles) and experiment data (solid circles) of our present work.

Fig. 9. TRS plots in the ($\beta_2$, $\gamma$) polar coordinate system for the positive-parity states in $^{80}$Rb with signature α=0 at ℏω=0.0 (upper left), 0.2 (upper right), 0.4 (lower left), and 0.6 MeV (lower right) corresponding to $I$~(0 − 15)ℏ. A prolate (oblate) shape corresponds to a triaxiality of γ=0°(-60°). The black dot represents the overall minimum in each panel. The contour lines are separated by 200 keV.

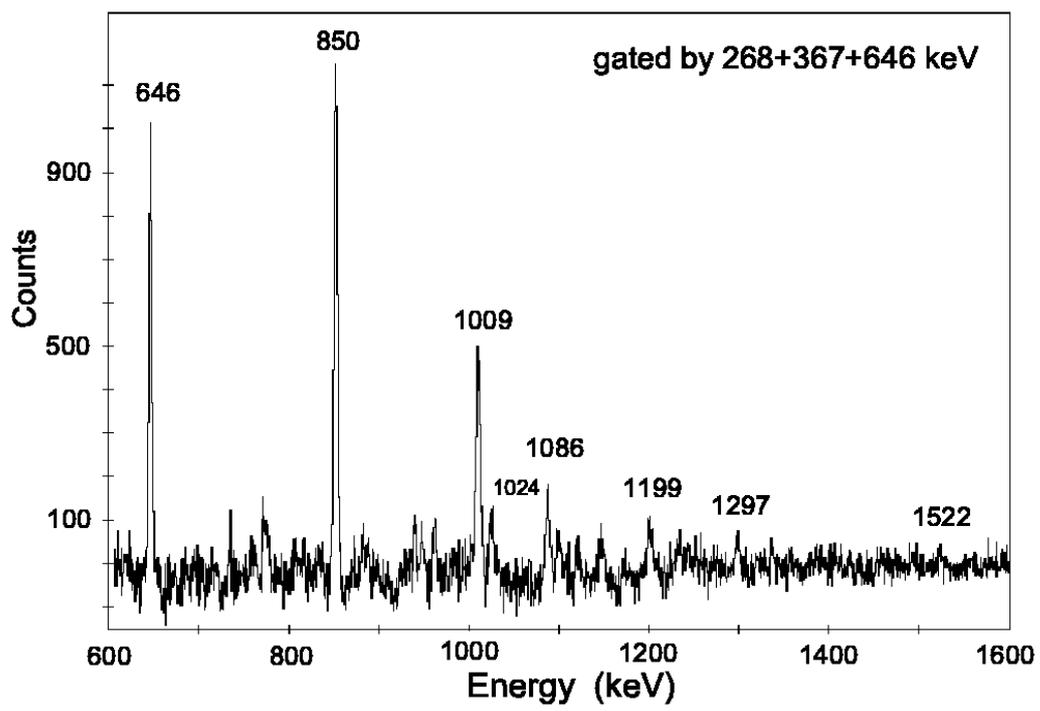

Fig. 1

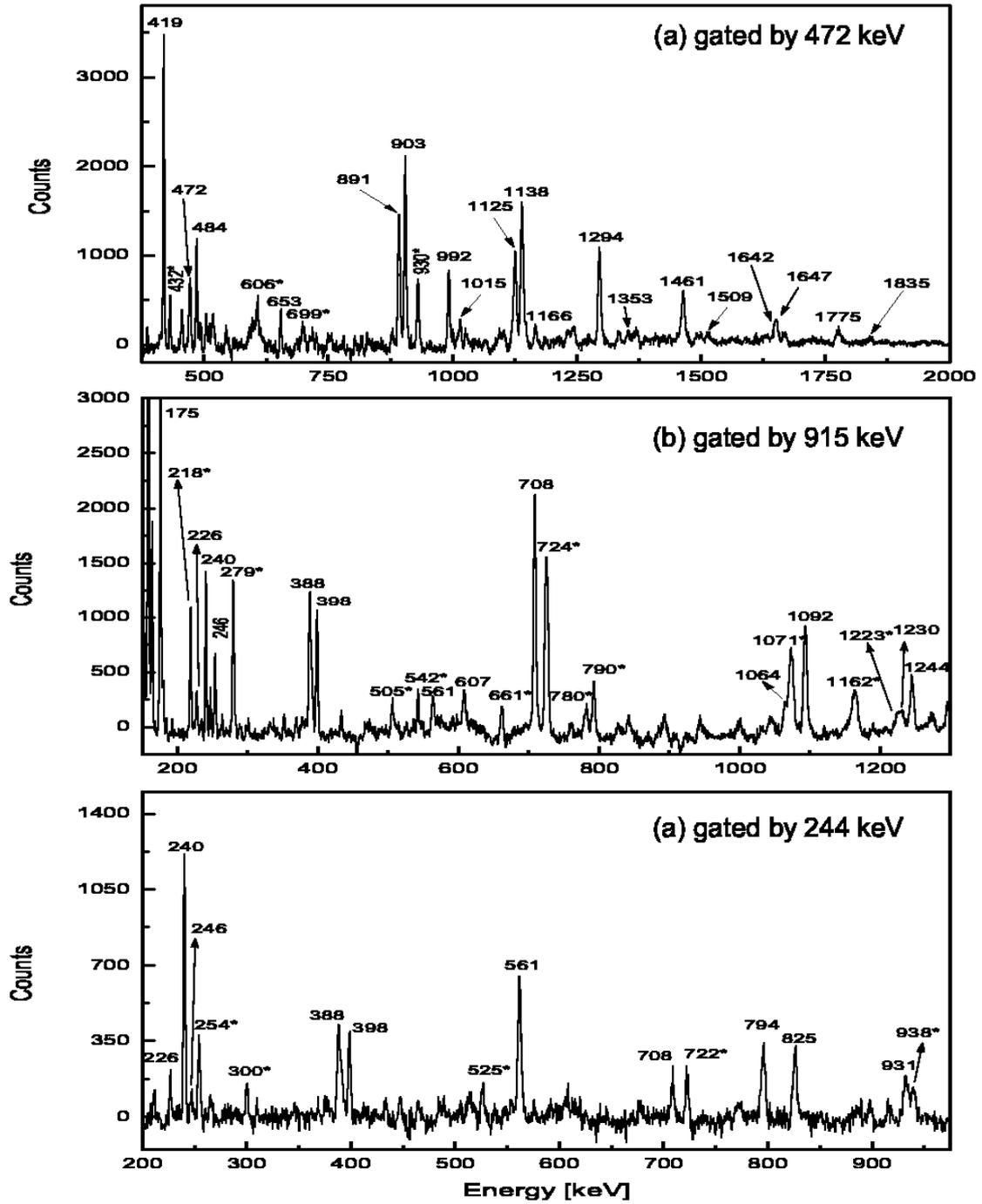

Fig. 2

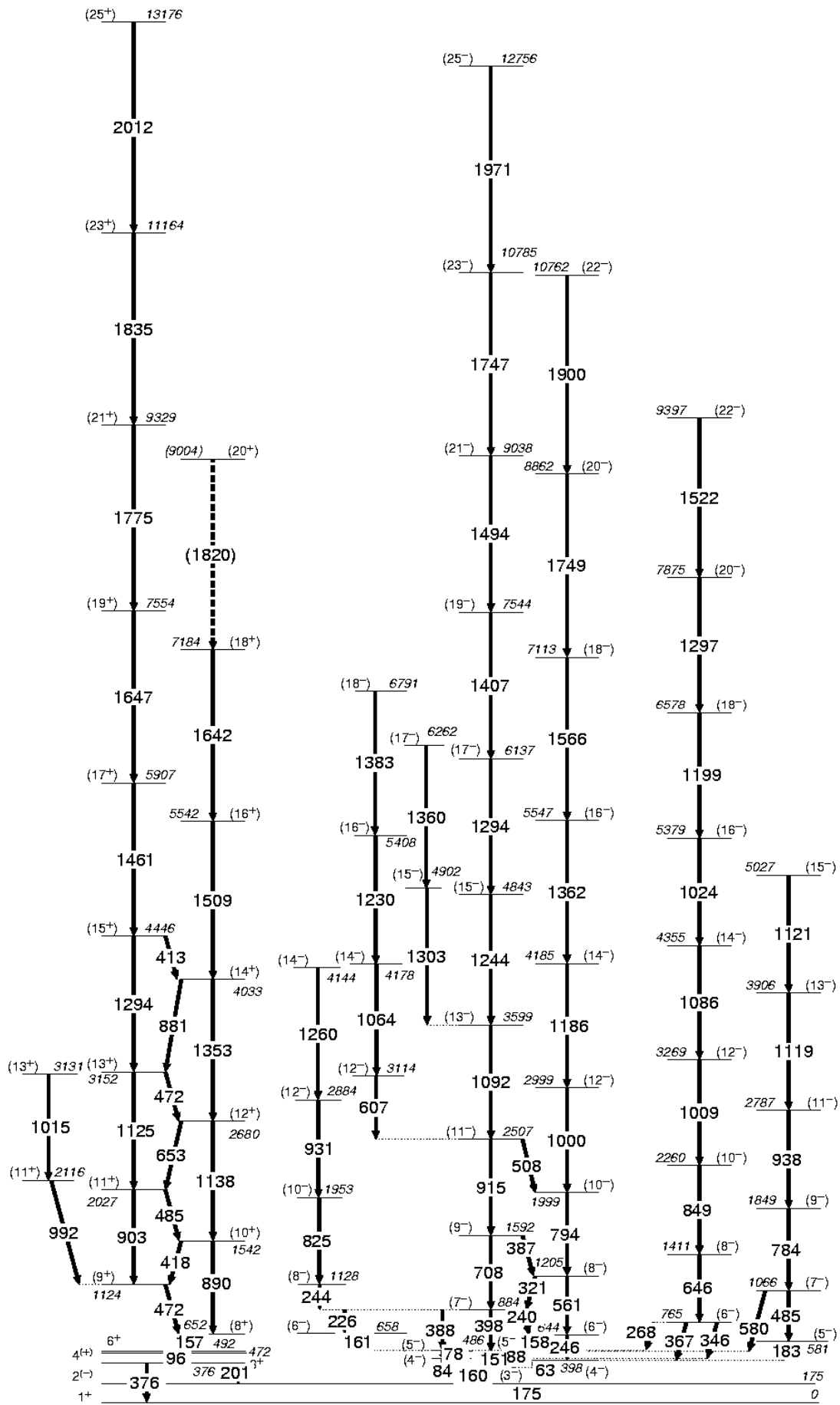

Fig. 3

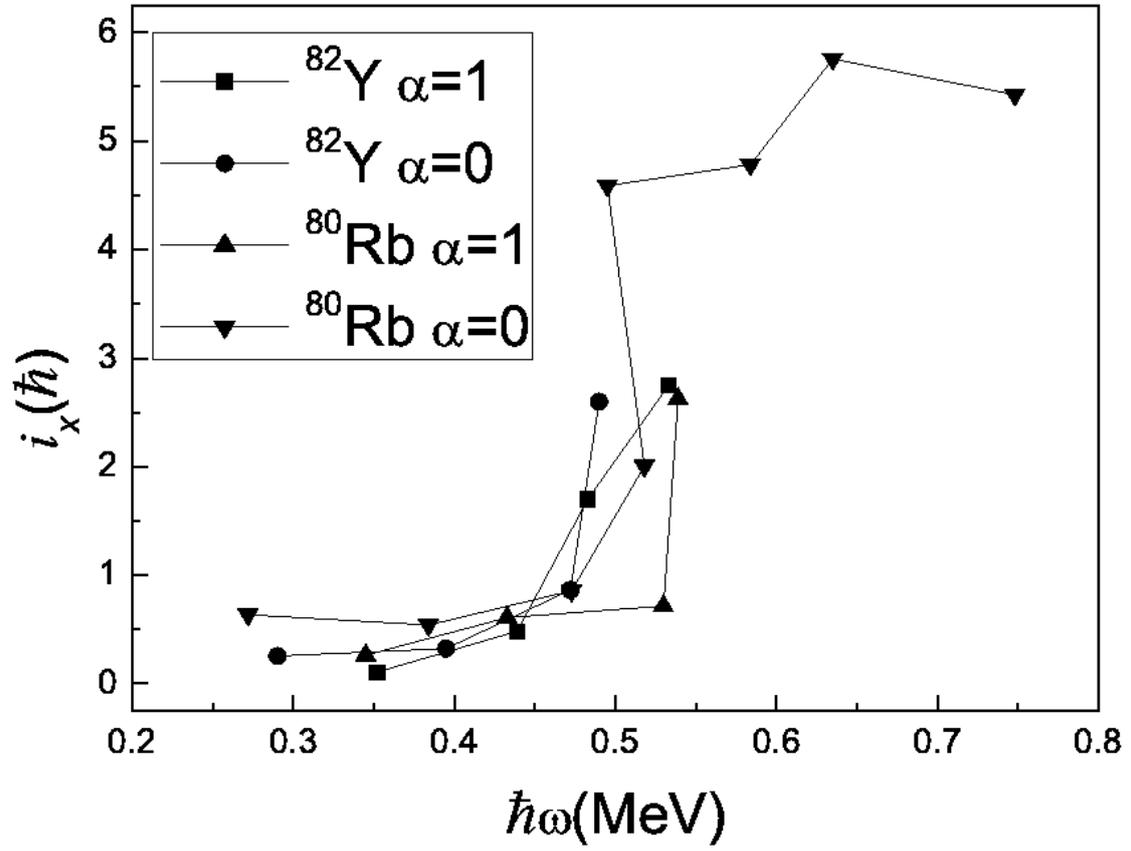

Fig. 4

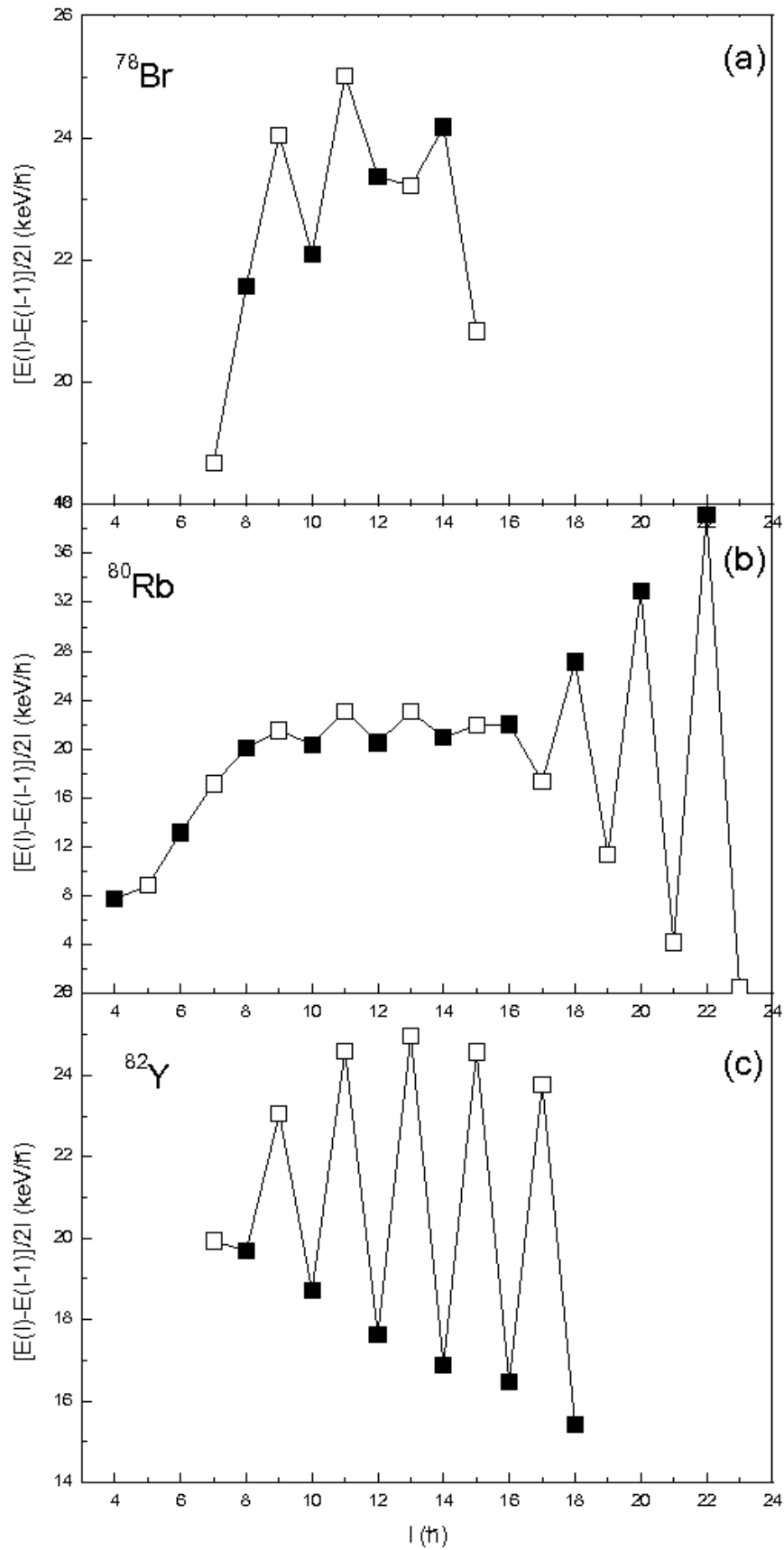

Fig. 5

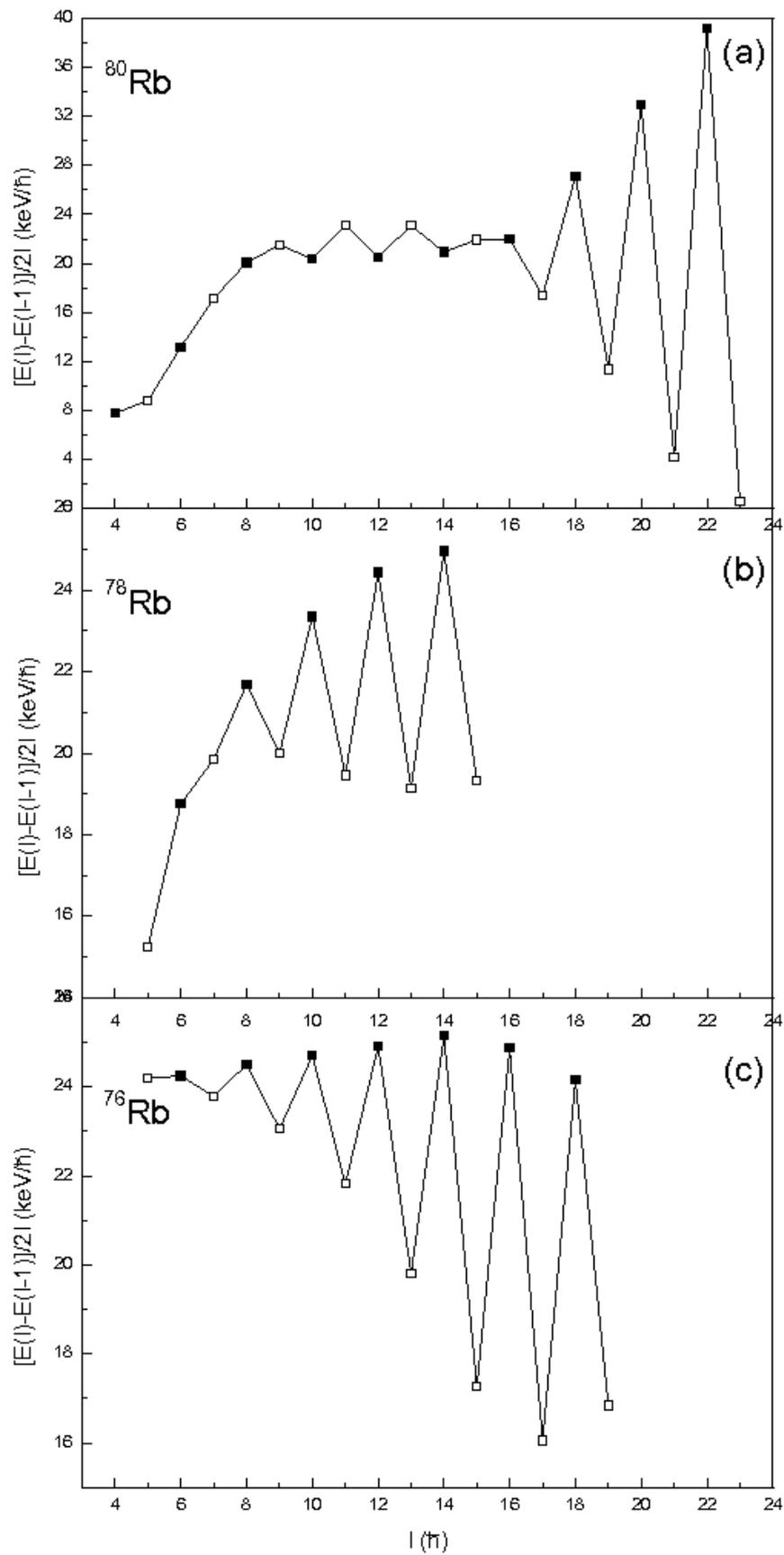

Fig. 6

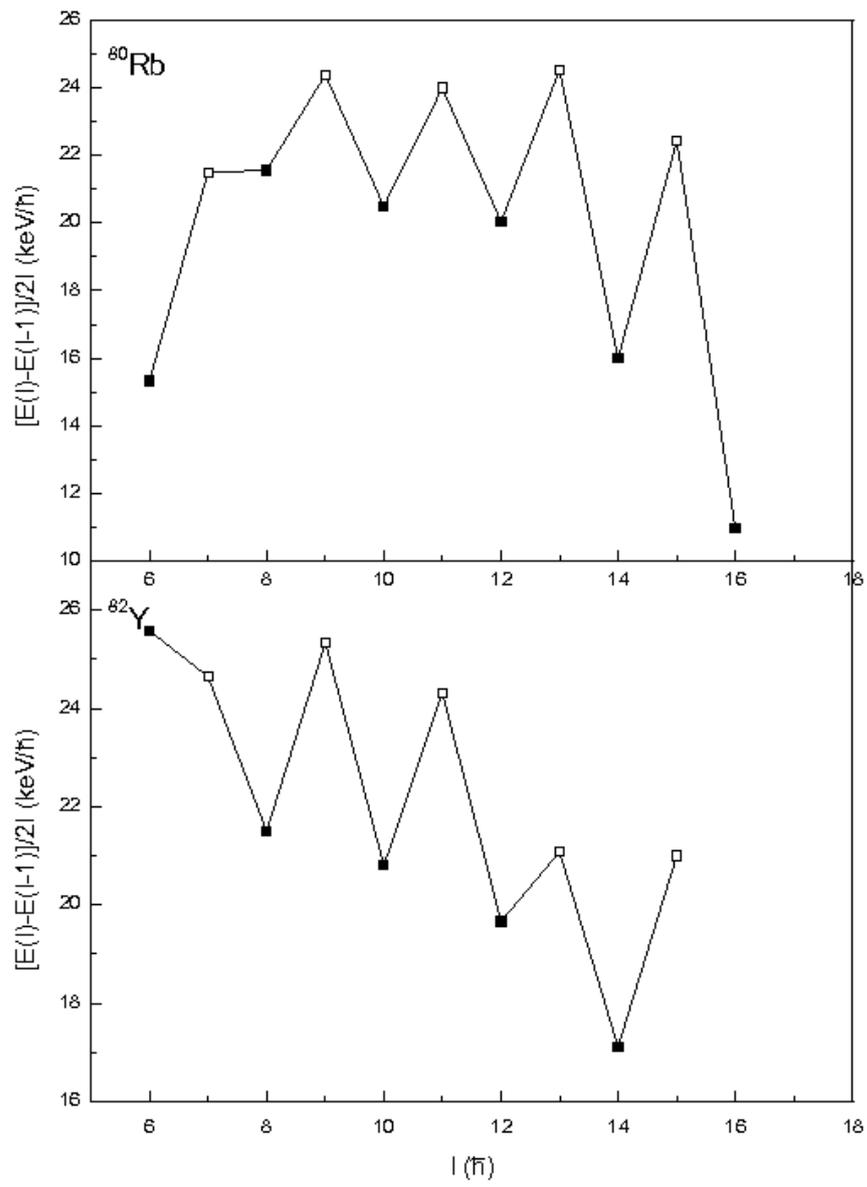

Fig. 7

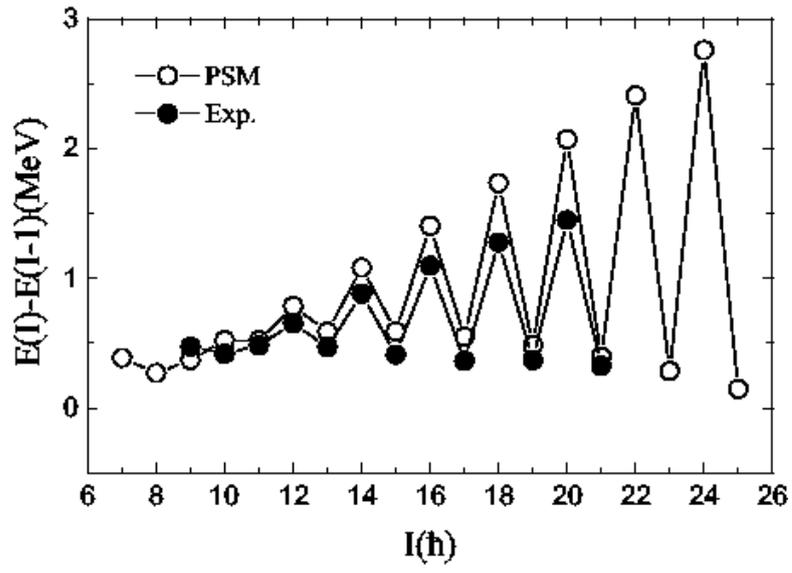

Fig. 8

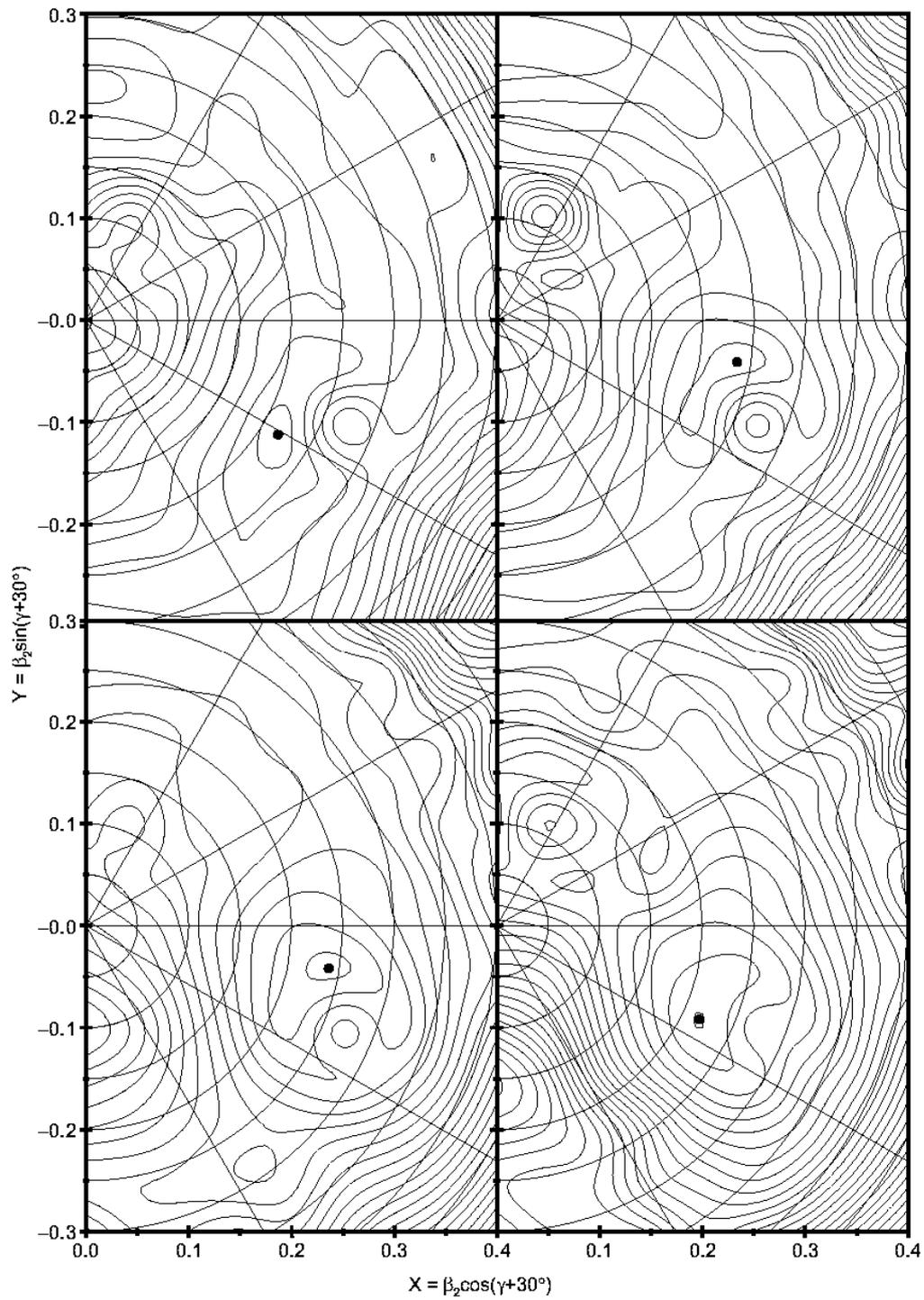

Fig. 9

TABLE I. The DCO ratios of strong γ rays in $^{80}$Rb deduced form the present experiment.

| $E_i^{\,a}$ | $E_\gamma^{\,b}$ | $E_\gamma^{gate\,c}$ | $I_i^{\pi\,d}$ | $I_f^{\pi\,e}$ | $R_{DCO}^{\,f}$ | $E_i^{\,a}$ | $E_\gamma^{\,b}$ | $E_\gamma^{gate\,c}$ | $I_i^{\pi\,d}$ | $I_f^{\pi\,e}$ | $R_{DCO}^{\,f}$ |
|---|---|---|---|---|---|---|---|---|---|---|---|
| (keV) | (keV) | (keV) | | | | (keV) | (keV) | (keV) | | | |
| 1542 | 418 | 472 | $(10^+)$ | $(9^+)$ | 0.95 | 6137 | 1294 | 175 | $(17^-)$ | $(15^-)$ | 1.20 |
| 2027 | 485 | 472 | $(11^+)$ | $(10^+)$ | 0.87 | 7544 | 1407 | 175 | $(19^-)$ | $(17^-)$ | 1.03 |
| 3152 | 472 | 472 | $(13^+)$ | $(12^+)$ | 0.80 | 9038 | 1494 | 175 | $(21^-)$ | $(19^-)$ | 0.80 |
| 2027 | 903 | 472 | $(11^+)$ | $(9^+)$ | 2.26 | 644 | 246 | 175 | $(6^-)$ | $(4^-)$ | 1.08 |
| 3152 | 1125 | 472 | $(13^+)$ | $(11^+)$ | 2.09 | 884 | 240 | 175 | $(7^-)$ | $(6^-)$ | 0.50 |
| 4446 | 1294 | 472 | $(15^+)$ | $(13^+)$ | 2.41 | 1205 | 561 | 175 | $(8^-)$ | $(9^-)$ | 1.08 |
| 5907 | 1461 | 472 | $(17^+)$ | $(15^+)$ | 2.28 | 1999 | 794 | 175 | $(10^-)$ | $(8^-)$ | 1.63 |
| 7554 | 1647 | 472 | $(19^+)$ | $(17^+)$ | 1.95 | 2999 | 1000 | 175 | $(12^-)$ | $(10^-)$ | 1.33 |
| 9329 | 1775 | 472 | $(21^+)$ | $(19^+)$ | 2.06 | 4185 | 1186 | 175 | $(14^-)$ | $(12^-)$ | 1.59 |
| 652 | 156 | 890 | $(8^+)$ | $6^+$ | 0.98 | 5547 | 1362 | 175 | $(16^-)$ | $(14^-)$ | 1.35 |
| 2680 | 1138 | 890 | $(12^+)$ | $(11^+)$ | 0.96 | 765 | 268 | 175 | $(6^-)$ | $(5^-)$ | 0.61 |
| 4033 | 1353 | 890 | $(14^+)$ | $(12^+)$ | 1.04 | 765 | 367 | 175 | $(6^-)$ | $(4^-)$ | 0.99 |
| 472 | 96 | 175 | $4^{(+)}$ | $3^+$ | 0.94 | 765 | 346 | 175 | $(6^-)$ | $(4^-)$ | 1.03 |
| 2709 | 160 | 175 | $(3^-)$ | $2^{(-)}$ | 0.95 | 1411 | 646 | 175 | $(8^-)$ | $(13^-)$ | 1.24 |
| 418 | 84 | 175 | $(4^-)$ | $(3^-)$ | 0.52 | 2260 | 849 | 175 | $(10^-)$ | $(14^-)$ | 1.26 |
| 496 | 78 | 175 | $(5^-)$ | $(4^-)$ | 0.63 | 3269 | 1009 | 175 | $(12^-)$ | $(14^-)$ | 1.70 |
| 398 | 63 | 175 | $(4^-)$ | $(3^-)$ | 0.60 | 4355 | 1086 | 175 | $(14^-)$ | $(15^-)$ | 1.67 |
| 486 | 88 | 175 | $(5^-)$ | $(4^-)$ | 0.50 | 581 | 183 | 175 | $(5^-)$ | $(4^-)$ | 0.52 |
| 884 | 398 | 175 | $(7^-)$ | $(5^-)$ | 0.94 | 1066 | 485 | 175 | $(7^-)$ | $(5^-)$ | 0.92 |
| 1592 | 708 | 175 | $(9^-)$ | $(7^-)$ | 1.08 | 1066 | 580 | 175 | $(7^-)$ | $(5^-)$ | 1.28 |
| 2507 | 915 | 175 | $(11^-)$ | $(9^-)$ | 1.33 | 1849 | 784 | 175 | $(9^-)$ | $(7^-)$ | 1.33 |
| 3599 | 1092 | 175 | $(13^-)$ | $(11^-)$ | 1.68 | 2787 | 938 | 175 | $(11^-)$ | $(9^-)$ | 1.47 |
| 4843 | 1244 | 175 | $(15^-)$ | $(13^-)$ | 1.26 | 3906 | 1119 | 175 | $(13^-)$ | $(11^-)$ | 1.46 |

[a]Energy of the initial state.
[b]Transition energy.
[c]Energy of the gating transition used for the determination of the DCO ratio.
[d]Spin and parity of the initial state.
[e]Spin and parity of the final state.
[f]DCO ratio.